\documentclass[referee]{aa}
\title{Dark matter, density perturbations and structure formation.}
\author{A. Del Popolo\inst{1,2,3}}
\offprints{A. Del Popolo - {\it E-mail adelpop@unibg.it}}
\institute{$^1$ Dipartimento di Matematica, Universit\`{a} Statale
di Bergamo,
via dei Caniana, 2 - I 24129 Bergamo, ITALY \\
$^2$ Feza G\"ursey Institute, P.O. Box 6 \c Cengelk\"oy, Istanbul, Turkey \\
$^3$ Bo$\breve{g}azi$\c{c}i University, Physics Department,
80815 Bebek, Istanbul, Turkey}

\begin{document}
\def\cl{\centerline}
\def\bk{\hfill\break}
\def\no{\noindent}

{\bf \huge Dark matter, density perturbations 
and structure formation.\\}

{\bf A. Del Popolo}\inst{1,2,3}\\
$^1$ Dipartimento di Matematica, Universit\`{a} Statale
di Bergamo,
via dei Caniana, 2 - I 24129 Bergamo, ITALY \\
$^2$ Feza G\"ursey Institute, P.O. Box 6 \c Cengelk\"oy, Istanbul, Turkey \\
$^3$ Bo$\breve{g}azi$\c{c}i University, Physics Department,
80815 Bebek, Istanbul, Turkey


\vspace{2.0cm}
\begin{abstract}
This paper provides a review of the variants of dark matter (CDM, HDM) which are thought to be fundamental components of the universe and their role in origin and evolution of structures. 
\end{abstract}

\begin{flushleft}
{\bf Introduction}
\end{flushleft}
The origin and evolution of large scale structure is today
the outstanding problem in cosmology. This is the most fundamental question
we can ask about the universe whose solution should help us to better understand problems
as the epoch of galaxy formation, the clustering in the galaxy
distribution, the amplitude and form of anisotropies in the microwave
background radiation. Several has been the approaches and models
trying to attack and solve this problem: no one has given a final answer. \\
The leading idea of all structure formation theories is that structures
was born from small perturbations in the otherwise uniform distribution
of matter in the early Universe, which is supposed to be, in great part,
dark (matter not detectable through light emission).

With the term Dark Matter cosmologists indicate an hypothetic material component of the universe which does not emit directly electromagnetic radiation (unless it decays in particles having this property (Sciama 1989, but also see Bowyer et al. 1999)).\\
Dark matter, cannot be revealed directly, but nevertheless it is necessary to postulate its existence in order to explain the discrepancies between the observed dynamical properties of galaxies and clusters of galaxies  
and the theoretical predictions based upon models of these objects assuming that the only matter present is the visible one.
If in the space were present a diffused material component having gravitational mass, but unable to emit electromagnetic radiation in significative quantity, this discrepancy could be eliminated (Turner 1991). 
The study of Dark Matter has as its finality the explanation of formation of galaxies and in general of cosmic structures. For this reason, in the last decades, the origin of cosmic structures has been ``framed" in models in which Dark Matter constitutes the skeleton of cosmic structures and supply the most part of the mass of which the same is made.\\
There are essentially two ways in which matter in the universe can be revealed: by means of radiation, by itself emitted, or by means of its gravitational interaction with baryonic matter which gives rise to cosmic structures. 
Electromagnetic radiation permits to reveal only baryonic matter. In the second case, we can only tell that we are in presence of matter that interacts by means of gravitation with the luminous mass in the universe. 
The original hypotheses on Dark Matter go back to measures performed by Oort (1932) of the surface density of matter in the galactic disk, which was obtained through the study of the stars motion in direction orthogonal to the galactic plane. The result obtained by Oort, which was after him named ``Oort Limit", gave a value of $ \rho = 0.15 M_{0} pc^{-3}$ for the mass density, and a mass, in the region studied, superior to that present in stars.
Nowadays, we know that the quoted discrepancy is due to the presence of HI in the solar neighborhood. Other studies (Zwicky 1933; Smith 1936) showed the existence of a noteworthy discrepancy between the virial mass of clusters (e.g. Coma Cluster) and the total mass contained in galaxies of the same clusters.
These and other researches from the thirties to now, have confirmed that a great part of the mass in the universe does not emit radiation that can be directly observed. \\

{\bf 1.1 Determination of $ \Omega $ and Dark Matter}\\

The simplest cosmological model that describes, in a sufficient coherent manner, the evolution of the universe, from $ 10^{-2}
s$ after the initial singularity to now, is the so called {\it Standard Cosmological Model} (or Hot Big Bang model). It is based upon the Friedmann-Robertson-Walker (FRW) metric, which is given by:
\begin{equation}
ds^{2} = c^{2} d t^{2} -a(t)^{2}\left[\frac{d r^{2} }{1 - k r^{2}}+
r^{2} (d \theta^{2} +sin{\theta}^{2} d \phi^{2} )\right]
\end{equation}
where c is the light velocity, a(t) a function of time, or a scale factor called ``expansion parameter", t is the time coordinate, r, $\theta $ and  $ \phi $ the comoving space coordinates. The evolution of the universe is described by the parameter a(t) and it is fundamentally connected to the value $\rho$ of the average density.\\
The equations that describe the dynamics of the universe are the Friedmann's equations (Friedman, 1924) that we are going to introduce in a while. These equations can be obtained starting from the equations of the gravitational field of Einstein (Einstein, A., 1915): 
\begin{equation}
R_{ik}-\frac{1}{2} g_{ik} R =-\frac{8 \pi G}{ c^{4}} T_{ik}
\end{equation}
where now, $ R_{ik} $ is a symmetric tensor, also known as Ricci tensor, which describes the geometric properties of space-time, $ g_{ik} $ is the metric tensor, R is the scalar curvature, $T_{ik}$ is the energy-momentum tensor.\\
These equations connect the properties of space-time to the mass-energy. In other terms they describe how space-time is modeled by mass. Combining Einstein equations to the FRW metric leads to the dynamic equations for the expansion parameter, a(t). These last are the Friedmann equations:
\begin{equation}
d( \rho a^{3} )= -p d( a^{3} )
\end{equation}
\begin{equation}
\frac{1}{a^{2}} \dot{a}^{2} +\frac{k}{a^{2}} = \frac{8 \pi G }{3} \rho
\end{equation}
\begin{equation}
2\frac{\ddot{a}}{a} + \frac{\dot{a}^{2}}{a^{2}} +\frac{k}{a^{2}} =
-8 \pi G \rho
\end{equation}
where p is the pressure of the fluid of which the universe is constituted, k is the curvature parameter and a(t) is the scale factor connecting proper distances $ { \bf r }  $  to the comoving ones $ {\bf x} $ through the relation 
$ {\bf r}=a(t) {\bf x} $. One of the components of the today universe are galaxies. If we assume that galaxies motion satisfy Weyl (Weyl, H., 1923) postulate, the velocity vector of a galaxy is given by 
$ u^{i} = (1,0,0,0) $, and then the system behaves as a system made of dust for which we have $p=0$.
Only two of the three Friedmann equations are independent, because the first connects density, $\rho$ to the expansion parameter a(t). The character of the solutions of these equations depends on the value of the curvature parameter, $k$, which is also determined by the initial conditions by means of Eq. 3. 
The solution to the equations now written shows that if $\rho$ is larger than $ \rho_{c} = \frac{3 H^{2} }{ 8 \pi G } = 1.88* 10^{-29} g/cm^{3} $ (critical density, which can be obtained from Friedmann equations putting $t=t_0$, $k=0$, and $ H= 100 km /s Mpc $), space-time  has a closed structure ($k=1$) and equations shows that the system go through a singularity in a finite time. This means that the universe has an expansion phase until it reaches a maximum expansion after which it recollapse. If $ \rho < \rho_{c} $, the expansion never stops and the universe is open $k=-1$ (the universe has a structure similar to that of an hyperboloid, in the two-dimensional case).
If finally, $ \rho = \rho_{c} $ the expansion is decelerated and has infinite duration in time, $k=0$, and the universe is flat (as a plane in the two-dimensional case). The concept discussed can be expressed using the parameter $ \Omega = \frac{\rho}{\rho_{c}} $. In this case, the condition $ \Omega = 1 $ corresponds to $k=0$, $ \Omega <1 $ corresponds to $k=-1$, and $ \Omega > 1 $ corresponds to $k=1$. \footnote{See next paragraphs for some items on cosmological models with non-zero cosmological constant.}\\
The value of $ \Omega $ can be calculated in several ways. The most common methods are the dynamical methods, in which the effects of gravity are used, and kinematics methods sensible to the evolution of the scale factor and to the space-time geometry. The results obtained for $\Omega$ with these different methods are summarized in the following.\\

{\it Dynamical methods}:\\

(a) Rotation curves: The contribution of spiral galaxies to the density in the universe is calculated by using their rotation curves and the third Kepler law. Using the last it is possible to obtain the mass of a spiral galaxy from the equation:
\begin{equation}
M(r) = v^{2} r /G  \label{eq:kep}
\end{equation}
where v is the velocity of a test particle at a distance r from the center and M(r) is the mass internal to the circular orbit of the particle. In order to determine the mass M is necessary to have knowledge of the term  $v^{2} $ in Eq. (\ref{eq:kep}) and this can be done from the study of the rotation curves through the 21 cm line of HI.  
Rotation curves of galaxies are characterized by a peak reached at distances of some Kpcs and a behavior typically flat for the regions at distance larger than that of the peak. A peculiarity is that the expected Keplerian fall is not observed. This result 
is consistent with extended haloes containing masses till 10 times the galactic mass observed in the optical (van Albada, T. S., Sancisi, 1986). The previous result is obtained 
assuming that the halo mass obtained with this method is distributed in a spherical region so that we can use Eq. (\ref{eq:kep}) and that we neglect the tidal interaction with the neighboring galaxies which tend to produce an expansion of the halo.
After M and the luminosity of a series of elliptical galaxies is determined, the contribution to the density of the universe is given by $ \rho = < \frac{M}{L} > \ell $ where $\ell$ is the luminosity per unit volume due to galaxies 
and can be obtained from the galactic luminosity function $ \phi(L) dL,  $ which describes the number of galaxies per $ Mpc^{3} $ and luminosity range $ L, L+dL $. The value that is usually assumed for $\ell$ is $ \ell= 2.4h 10^{8} L_{bo} Mpc^{-3} $. The arguments used lead to a value of $\Omega_{g}$ for the luminous parts of spiral galaxies of $\Omega_{g} \leq 0.01$, while for haloes $ \Omega_{h} \geq 0.03 -0.1 $. The result shows that the halo mass is noteworthy larger than the galactic mass observable in the optical (Peebles,P.J.E.,1971).\\

{(b) \it Virial theorem}:\\

In the case of non spiral galaxies and clusters, the mass can be obtained using the virial theorem $ 2 T + V = 0 $, with 
\begin{equation}
T \cong \frac{3}{2} M <v_{r}^{2}> 
\end{equation}
where $ < v_{r}^{2} > $ is the velocity dispersion along the line of sight. After getting the value of M of the cluster by means of the virial theorem one determines L by means of observations. Given M and L, the value of $ \Omega $ for clusters is obtained similarly to the case of spiral galaxies. Usual values obtained for $\Omega$ are $ \Omega = 0.1-0.3 $ (Peebles, P.J.E., 1971). A problem of the quoted method is that in general the results obtained are the right one only for virialized, spherically symmetric clusters. In general, clusters are not virialized objects: even Coma clusters seems to have a central core constituted by more than one blob of mass (Henry, J.P.,Briel, U.G.,1992).  \\

{(c) \it Peculiar velocities}: \\

The velocity of a galaxy can be written as:
\begin{equation}
{\bf V}_{g} =H{\bf r} + \delta{\bf v}
\end{equation}
where $H$ is Hubble constant. The previous equation shows that the motion of a galaxy is constituted by two components:
the velocity of the galaxy due to the Hubble flow and a peculiar velocity $\delta{\bf v}$, which describes the motion of the galaxy with respect to the background. In the linear regime, as we see in a while, we find that on average on a scale of length $\lambda$, it is:
\begin{equation}
\frac{\delta v}{c} \approx \Omega^{0.6}
\frac{\lambda}{H_{0}^{-1}}\frac{\delta \rho}{\rho}
\end{equation}
(Kolb e Turner 1990).
Then given the overdensity $ \frac{\delta \rho}{\rho} $ on scale $ \lambda $ and $ \delta v $, it is possible to obtain $\Omega$. The overdensity $ \frac{\delta \rho}{\rho} $ can be obtained from the overdensity of galaxies $ \frac{\delta n_{g}}{n_{g}}$ using the relation $ \frac{\delta \rho}{\rho} = \frac{\delta n_{g} }{n_{g}} b^{-1} $
with $ 1<b<3 $. Using IRAS catalog in order to obtain the overdensity in galaxies one finds $ \Omega \cong 1 $ ( M.S.Turner ). The values of $\Omega$ obtained using the method of peculiar velocity assume that the peculiar velocity fields describe in an accurate way, the inhomogeneity in the distribution of underlying mass. 
We should note that the peculiar velocity method has some difficulties. In general, in order to obtain these last it is necessary to determine the redshift and the distance of galaxies and by using these data it is possible to obtain the peculiar velocity:   
\begin{equation}
v_{pec} = z c - H_{0} d
\end{equation}
It is evident that there are problems in measuring the distance d, problems connected to difficulties in finding  trustable indicators of distance. Moreover the peculiar velocity can be determined only along the line of sight. \\

{(d) \it Kinematic methods}:\\

These methods are based upon the use of relations between physical quantities dependent on cosmological parameters. An example of those relations is the relation luminosity distance-redshift:
\begin{equation}
H_{0} d_{L} = z+\frac{1}{2}(1-q_{0} )z^{2}
\end{equation}
where $ H_{0} $ is Hubble constant nowadays, z is the redshift, $ d_{L} = \frac{4 \pi L}{F} $ the luminosity distance, L the absolute luminosity, and F the flux. 
By means of the relations luminosity-redshift, angle-redshift, number of objects-redshift, it is possible to determine the parameter of deceleration $ q_{0} = -\frac{\ddot{a_{0}}}{H_{0}^{2} a_{0}} $ ($ a_{0} $ and 
$ H_{0}=\frac{\dot{a_{0}}}{a_{0}}=100 h km /Mpc s $ are the scale factor and the Hubble constant, nowadays). 
At the same time $ q_{0} $ is connected to $ \Omega $ by means of $ q_{0} = \frac{\Omega}{2} $, in a matter universe.
One of the first test used, the luminosity distance-redshift has several problems due to effects of the evolution of sources. Uncertainties in the knowledge of the effects of galactic evolution on the intrinsic luminosity of the same has not, in the past, permitted to find definitive values of  $ q_{0} $. For this reason, Loh \& Spillar (1986) introduced another kinematic test: number of galaxies-redshift. This test is based on the count of the number of galaxies in a comoving element of galaxies, defined by the surface $ d \Omega $ and the redshift $d z$. 
This number depends on $ q_{0} $. Nevertheless the effects of evolution of sources influences on the results of the test, it is more sensible to the evolution of number of sources than to the evolution of luminosity, on which there is not an accepted theory. Results gives high values of $\Omega$ ($\Omega =0.9_{-0.5}^{+0.7} $) (Kolb e Turner 1990). 
\footnote{Among kinematics methods we should mention SNeIa which played a key role in the last few years (Valageas 1999).}\\

{(e) \it Primordial nucleosynthesis}:\\
The theory of primordial nucleosynthesis, proposed in 1946 by Gamov, assumes that the light elements till $ Li_{7} $ are generated after big bang and that heavier elements originate from nuclear reactions inside stars. The values obtained for the abundances depends on some parameters like: $ \eta $, the value of the ratio baryons-photons, nowadays; $ N_{\nu} $, the number of neutrinos species; $ T_{CMBR} $, the temperature of Cosmic Microwave Background Radiation. 
The theory of primordial nucleosynthesis permits to give limits to $\Omega_b$ (b stands for baryons). With a ratio baryons-photons $ 3 *10^{-10} \leq \eta \leq 5 *10^{-10} $, and a value of $ N_{\nu} \leq 4 $ for the neutrinos species, $ T_{CMBR}= 2.736 \pm 0.01 K $, is found $ 0.011\leq \Omega_{b} \leq 0.12 $ (Kolb e Turner 1990 ).\\


In the following, we summarize the results of some more recent results. 

Dekel et al 1996 used several methods to obtain the value of $\Omega$.
According to their classification, we divide the methods into the following four classes:

\leftskip=1.0 true cm

\leftskip=1.0 true cm
\noindent
{\hskip -0.5 true cm}
$\bullet\ ${\it Global measures}. Based on properties of space-time
that constrain combinations of $\Omega_m$ and the
other cosmological parameters ($\Lambda$, $H_0$, $t_0$).

\leftskip=1.0 true cm
\noindent
{\hskip -0.5 true cm}
$\bullet\ ${\it Virialized Systems}. Methods based on nonlinear dynamics 
within galaxies and clusters on comoving scales $1-10 h^{-1} Mpc$.

\leftskip=1.0 true cm
\noindent
{\hskip -0.5 true cm}
$\bullet\ ${\it Large-scale structure}. Measurements based on 
mildly-nonlinear gravitational dynamics of fluctuations on scales 
$10-100 h^{-1} Mpc$ of superclusters and voids, in particular {\it cosmic flows}.

\leftskip=1.0 true cm
\noindent
{\hskip -0.50 true cm}
$\bullet\ ${\it Growth rate of fluctuations}. Comparisons of present day
structure with fluctuations at the last scattering of the cosmic microwave 
background (CMB) or with high redshift objects of the young universe.

\leftskip=0.0 true cm

The methods and current estimates are summarized in Table 1.
The estimates based on virialized objects typically yield low values of
$\Omega_m \sim 0.2-0.3$. The global measures, large-scale structure and cosmic
flows typically indicate higher values of $\Omega_m \sim 0.4-1$.
\newpage
 \bigskip\bigskip

\def \srule {
        \vskip 0.4\baselineskip
        \hrule height.7pt
        \vskip 0.3\baselineskip }
\def \drule {
        \vskip 0.4\baselineskip
        \hrule height.7pt
        \vskip2pt
        \hrule height.7pt
        \vskip 0.3\baselineskip }
 
\cl{\bf Table 1: Estimates of {$\Omega_m$}}
{\bf Table 1: Estimates of {$\Omega_m$}}
\smallskip
 
\vbox {
\drule
\halign to \hsize {
#\quad\hfil&#\quad\hfil&#\hfil\cr
 
&&\cr
Global Measures
&Inflation, Occam
&$\Omega_m+\Omega_{\Lambda}=1$ \ ($\Omega_m=1$, $\Omega_{\Lambda}=0$)          \cr
&Lumimosity distance SNIa
&$-0.3<\Omega_m-\Omega_{\Lambda}<2.5$ (90\%) \
\cr
&&Flat $\Omega_m > 0.49$ (95\%)     \cr
&Lens Counts
&Flat $\Omega_m > 0.34$ (95\%)                        \cr
&CMB Peak
&$\Omega_m + \Omega_{\Lambda} < 1.5$ (95\%) \cr
&
&$\Omega_m + \Omega_{\Lambda} > 0.3$ \ (likely $\sim 0.7$)         \cr
&$H_0 t_0$
&$ \Omega_m -0.7 \Omega_{\Lambda} < 1.3$ \ (likely $\leq 0$)   \cr
 
&&\cr
Virialized Objects
%
&$(M/L){\cal L}$
&$\Omega_m \sim 0.25$  ($0.1-1.0$)                        \cr
&Baryon fraction
&$\Omega_m h_{65}^{1/2} \sim 0.3-0.5$ (low$-$high $\Omega_b$) \cr
&Cosmic Virial Th.
&Point mass $\Omega_m \!\sim\! 0.2$ (halos $\!\rightarrow\!1$)      \cr
&Local Group
&Point mass $\Omega_m \!\sim\! 0.15$ (halos $\!\rightarrow\!0.7$)  \cr
 
&&\cr
Large-Scale; Flows
&Peculiar velocities
&$\Omega_m > 0.3$ (95\%) \cr
&
&$\Omega_m^{0.6} \sigma_8^a =0.8 \pm 0.2$ \ ($\beta_I^b \simeq 1.05^c$)\cr
&Redshift Distortions
&$\beta_I \sim 0.5-1.2$  \cr
&Velocity vs Density
&$\beta_I \!\sim 0.5-1.2$ (scale dependent) \cr
&&$\beta_o \!\sim 0.4-0.95$   \cr
&Cluster Abundance
&$\Omega_m^{0.6} \sigma_8 \simeq 0.5-0.6$ \ ($\beta_I \simeq 0.7-0.8^c$)\cr
 
&&\cr
Fluctuation Growth
%
&Cluster Morphology
&$\Omega_m > 0.2$ (?)         \cr
%
%
&Galaxy Formation
&(?) \cr
&$P_k(\rho)$ vs $C_l$
&CDM $n=1$ $b=1$: $\Omega_m h_{65} \sim 0.3$  \cr
&$P_k(v)$ vs $C_l$
&CDM flat: $\Omega_m h_{65} n^2 \simeq 0.7 \pm 0.1$  \cr
&& \cr
\noalign {\srule}
&&\cr
}
}
\no$^a$ See Dekel et al. 1996 for a discussion of the table. 
$\sigma_8$ is the {\it rms} mass density fluctuation in a top-hat
    sphere of radius $8 h^{-1} Mpc$. \bk \\
\no$^b$ $\beta \equiv \Omega^{0.6}/b$,\ \ ~$b_I$ for IRAS galaxies,
                         ~ $b_o$ for optical galaxies. \bk \\ 
\no$^c$ $b_o/b_I \simeq 1.3$,\ \ ~$b_o\simeq 1/\sigma_8$.

Neta Bahcall et al. 1997,
showed that the  evolution of the number density of rich clusters of galaxies
breaks the degeneracy between $\Omega$ (the mass density
ratio of the universe) and
$\sigma_{8}$ (the normalization of the power spectrum), $\sigma_{8} \:
\Omega^{0.5} \simeq 0.5$, that
follows from the observed present-day abundance of rich clusters.  The
evolution of high-mass
(Coma-like) clusters is strong in $\Omega = 1$, low-$\sigma_{8}$ models (such as
the standard biased CDM model with $\sigma_{8} \simeq 0.5$), where the
number density of clusters
decreases by a factor of $\sim 10^{3}$ from $z = 0$ to $z \simeq 0.5$; the
same clusters show only mild evolution in low-$\Omega$, high-$\sigma_{8}$ models,
where the decrease is
a factor of $\sim 10$.
This diagnostic provides a most powerful constraint on $\Omega$.
Using observations of clusters
to $z \simeq 0.5-1$, the authors found  
only mild evolution in the observed cluster abundance, and 
$\Omega = 0.3 \pm 0.1$ and $\sigma_{8} = 0.85 \pm 0.15$
(for $\Lambda = 0$ models;
for $\Omega + \Lambda = 1$ models, $\Omega = 0.34 \pm 0.13$). 

Ferreira et al. 1999, proposed an
alternative method to estimate $v_{12}$ directly from peculiar velocity
samples, which contain redshift-independent distances as well as galaxy
redshifts.
In contrast to other dynamical
measures which determine $\beta\equiv\Omega^{0.6}\sigma_8$, this method
can provide an estimate of $\Omega^{0.6}\sigma_8^2$
for a range of $\sigma_8$
where $\Omega$ is the cosmological density parameter, while $\sigma_8$
is the standard normalization for the power spectrum of density
fluctuations. 

Melchiorri et al. 1999, used the angular power spectrum of the Cosmic Microwave Background,
measured during the North American test flight of the BOOMERANG experiment,
to constrain the geometry of the universe. Within the class
of Cold Dark Matter models, they find 
that the overall fractional energy density of the universe, $\Omega$, is
constrained to be $0.85 \le \Omega \le 1.25$ at the $68\%$ confidence level.

Branchini et al. 1999, compared 
the density and velocity fields as extracted from the Abell/ACO
clusters to the corresponding fields recovered by 
the POTENT method from the Mark~III peculiar velocities of galaxies.
Quantitative comparisons within a volume containing $\sim\!12$ 
independent samples yield $\beta_c \equiv \Omega^{0.6}/b_c=0.22\pm0.08$, 
where $b_c$ is the cluster biasing parameter at $15 h^{-1} Mpc$. If $b_c
\sim 4.5$, as indicated by the cluster correlation function, their
result is consistent with $\Omega \sim 1$.

{(f) \it Inflation}:\\

It is widely supposed that the very early universe experienced an era of
inflation (see Guth 1981, Linde 1990, Kolb \& Turner 1990). By `inflation' one means 
that the scale factor has positive
acceleration, $\ddot a>0$, corresponding to repulsive gravity and $3p<-\rho$.
During inflation $aH=\dot a$ is increasing, so that comoving scales are
leaving the horizon (Hubble distance) rather than entering it, and it is
supposed that at the beginning of inflation the observable universe was well
within the horizon.

The inflationary hypothesis is attractive because it holds out the possibility
of calculating cosmological quantities, given the Lagrangian describing the
fundamental interactions. The Standard Model, describing the interactions up
to energies of order $1 TeV$, is not viable in this context because it does
not permit inflation, but this should not be regarded as a serious setback
because it is universally agreed that the Standard Model will require
modification at higher energy scales, for reasons that have nothing to do with
cosmology. The nature of the required extension is not yet known, though it is
conceivable that it could become known in the foreseeable future (Weinberg
1993). But even without a specific model of the interactions (ie., a specific
Lagrangian), the inflationary hypothesis can still offer guidance about what
to expect in cosmology. More dramatically, one can turn around the
theory-to-observation sequence, to rule out otherwise reasonable models.
The importance of inflation is connected to:\\
a) the origin of density perturbations, which 
could originate during inflation as quantum fluctuations, which become
classical as they leave the horizon and remain so on re-entry. The original
quantum fluctuations are of exactly the same type as those of the
electromagnetic field, which give rise to the experimentally observed Casimir
effect.\\
b) One of the most dramatic and
simple effects is that there is no fine-tuning of the initial value of the
density parameter $\Omega=8\pi\rho/3 m_{Pl}^2 H^2$. From the
Friedmann equation, $\Omega$ is given by
\begin{equation} 
\Omega-1=(\frac{K}{aH})^2 \label{51} 
\end{equation}
Its present value $\Omega_0$ is certainly within an order of magnitude of 1,
and in the absence of an inflationary era $\Omega$ becomes ever smaller as one
goes back in time, implying an initial fine tuning. In contrast, if there is
an inflationary era beginning when the observable universe is within the
horizon, Eq. (\ref{51}) implies that $\Omega_0$ will be of order 1, provided only
that the same is true of $\Omega$ at the beginning of inflation. A value of
$\Omega_0$ extremely close to 1 is the most natural, though it is not
mandatory.\footnote {An
argument has been given for $\Omega_0$ very close to 1 on the basis of effects
on the cmb anisotropy from regions far outside the observable universe (Turner
1991), but it is not valid as it stands because it ignores spatial curvature.}\\
c) Another effect of inflation is that it can eliminate particles and topological
defects which would otherwise be present. Anything produced before inflation
is diluted away, and after inflation there is a maximum temperature (the
`reheat' temperature) which is not high enough to produce all the particles
and defects that might otherwise be present. As we shall remark later, this
mechanism can remove desirable, as well as undesirable, objects.\\
d) 
The most dramatic effect of inflation is that it may offer a way of
understanding the homogeneity and isotropy of the universe, or at any rate of
significant regions of it. We have nothing to say about this complex issue in
its full generality,  but a more modest version of it is our central concern.
In this version, one begins the discussion at some early stage of inflation,
when the universe is supposed already to be {\it approximately} homogeneous
and isotropic.
 One then argues that in that case, scales far inside the
horizon must be {\it absolutely} homogeneous and isotropic, except for the
effect of vacuum fluctuations in the fields. Finally, one shows that after
they leave the horizon, such fluctuations can become the classical
perturbations that one deals with in cosmological perturbation theory. This
possibility was first pointed out for gravitational waves by Starobinsky
(1979) and for density perturbations by several people (Guth \& Pi 1982;
Hawking 1982; Starobinsky 1982). As we shall go to some trouble to demonstrate,
the vacuum fluctuations can be evaluated unambiguously once an inflationary
model is specified.

{(g) \it Scalar field inflation}:\\

Two mechanisms for inflation have been proposed. The simplest one
(Guth 1981) invokes a scalar field, termed the inflaton field.
An alternative (Starobinsky 1980) is to invoke a modification of
Einstein gravity, and
combinations of the two mechanisms have also been proposed. During inflation
however, the proposed modifications of gravity
can be abolished by redefining the spacetime metric tensor, so that one
recovers the scalar field case. We focus on it for the moment,
but modified gravity models
will be included later in our survey of
specific models.

In comoving coordinates a homogeneous scalar field
 $\phi$ with minimal coupling to gravity has the
equation of motion
\begin{equation} 
\ddot \phi+3 H\dot \phi +V^\prime (\phi) =0 \label{52} 
\end{equation}
Its energy density and pressure are 
\begin{eqnarray}
\rho&=& V+\frac12\dot\phi^2\\
p&=&-V +\frac12\dot\phi^2 
\end{eqnarray}
If such a field dominates $\rho$ and $p$, the inflationary condition
$3p<\rho$ is achieved provided that the field rolls sufficiently slowly,\begin{equation} \dot\phi^2<V \end{equation}

Practically all of the usually considered models of inflation satisfy three
conditions. First, the motion of the field is overdamped, so that the `force'
$V^\prime $ balances the `friction term' $3H\dot\phi$,
\begin{equation}
\dot{\phi} \simeq -\frac{1}{3H} V' \label{56}
\end{equation}
Second,
\begin{equation}
\epsilon \equiv \frac{m_{Pl}^2}{16\pi}
\left( \frac{V'}{V} \right)^2 \ll 1 \label{57} 
\end{equation}
which means that the inflationary requirement $\dot\phi^2<V$ is well
satisfied and
\begin{equation} H^2 \simeq \frac13 \frac{8\pi}{m_{pl}^2} V \label{57a} \end{equation}
These two conditions imply that
$H$ is slowly varying, and that the scale factor increases more or less
exponentially,
\begin{equation} a\propto e^{Ht} \label{58b} \end{equation}
The third condition that is usually satisfied is
\begin{equation} |\eta|\ll1 \label{58} 
\end{equation}
where
\begin{equation} 
\eta \equiv \frac{m_{Pl}^2}{8\pi} \frac{V''}{V} 
\end{equation}
It can be `derived' from the other two by differentiating
the approximation Eq. (\ref{56}) for $\dot\phi$ and noting that consistency with the
exact expression  Eq. (\ref{52}) requires $\ddot \phi\ll V^\prime $ is satisfied.
However
there is no logical necessity for the derivative of an approximation to be
itself a valid approximation, so this third condition is logically independent
of the others. Conditions involving higher derivatives of $V$ could be
`derived' by further differentiation, with the same caveat, but the two that
we have given, involving only the first and second derivatives, are the ones
needed to obtain the usual predictions about inflationary perturbations. The
term `slow-roll inflation' is generally taken to denote a model in which they
are satisfied and we are adopting that nomenclature here. Practically all of
the usually considered models of inflation satisfy the slow-roll conditions
more or less well.

It should be noted that the first slow-roll condition is on a quite different
footing from the other two, being a statement about the {\em solution} of the
field equation as opposed to a statement about the potential that
defines this equation. What we are saying is that in the usually
considered models one can show
that the first condition is an attractor solution, in a regime typically
characterized by the other two conditions, and that moreover reasonable initial
conditions on $\phi$ will ensure that this solution is achieved well before
the observable universe leaves the horizon.

{(h) \it Conclusions}:\\
We have seen the possible values of $\Omega$ using different methods. We have to add that Cosmologists are ``attracted" 
by a value of $\Omega_0=1$. This value of $\Omega$ is requested by inflationary theory.
The previous data lead us to the following hypotheses:\\
i) $ \Omega_{0} < 0.12 $; in this case one can suppose that the universe is fundamentally made of baryonic matter (black holes; Jupiters; white dwarfs).\\
ii) $\Omega_{0} > 0.12$; in this case in order to have a flat universe, it is necessary a non-baryonic component. 
$\Omega_b=1$ is excluded by several reasons (see Efstathiou 1990 Kolb e Turner 1990).
The remaining possibilities are:\\
1) existence of a smooth component with $ \Omega = 0.8 $. \\
The test of a smooth component can be done with kinematic methods. \\ 
2) Existence of a cosmological term, absolutely smooth to whom correspond an energy density
$ \rho_{vac} =\frac{ \Lambda }{8 \pi G } $. \\
3) existence of non-baryonic matter: the universe is fundamentally done of particles (neutrinos, WIMPS (Weakly Interacting Massive Particles)). \\
4) A combination of 2) and 3).\\
Before going one, I want to recall that some authors (Finzi 1963; Milgrom 1983; Sanders 1984 ) have assumed that we have a scant knowledge of physical laws. Sanders assumes that the gravitational potential changes with distance and in particular the gravitational constant has a different value at large distances. Milgrom assumes that the Newton law of gravitation is not valid when the gradient of the potential is small. In this case, the problem of the dynamics of clusters of galaxies is solved without introducing Dark Matter. In any case, the quoted assumptions have no general theory that can justify them.\\

{\bf 1.2 Dark matter in particles.\\}

We know that if $ \Omega=1 $, dark matter cannot be constituted exclusive of baryonic matter. The most widespread hypothesis is that dark matter is in form of particles. Several candidates exist: neutrinos, axions, neutralinos, photinos, gravitinos, etc. Interesting particles are usually grouped into three families:\\
HDM (Hot Dark Matter), CDM (Cold Dark Matter) and WDM (Warm Dark Matter). In order to understand this classification it is necessary to go back to the early phases of universe evolution. The history of the universe is characterized by long phases of local thermodynamic equilibrium (LTE) and by ``deviation" by it: nucleosynthesys, bariogenesys, decoupling of species, etc. In the early universe, were present the particles that we know today and other particles predicted theoretically, but that have not been observed. Massive particles preserved the thermodynamic equilibrium concentration 
until the rate, $ \Gamma $, of reactions and interactions that produced that concentration was larger than the expansion rate of the universe, H. When the condition $ \Gamma > H $ was no longer satisfied the reactions stopped and the abundance of the considered species remained constant at the value it had at time of freez-out, $ T_{f} $, time at which $ \Gamma = H $. If we indicate with $ Y=\frac{n}{s} $ the number of particles per unit comoving volume and we remember that n is the number density of species and s the entropy density, we obtain a contribution of the species to the actual density of the universe as $ \Omega h^{2} =0.28 Y(T_{f} )(\frac{m}{ev}) $ (Kolb e Turner 1990). At time $ T_{f} $, particles could be relativistic or non-relativistic. Relativistic particles are today indicated with the term hot cosmic relics, HDM, while non-relativistic particles are named cold cosmic relics, CDM. There is an intermediate case, that of warm relics, WDM.\\
An example of HDM are massive neutrinos. Possible masses for these neutrinos are: 
\begin{equation}
25 ev \leq m_{\nu} \leq 100 ev 
\end{equation}
( Sciama, D.W., 1984) and 
\begin{equation}
m_{\nu} \geq \left\{\begin{array}{ll}
4.9-1.3 Gev & \mbox{for Maiorana's neutrinos}\\
1.3-4.2 Gev & \mbox{for Dirac's neutrinos}
\end{array}\right.
\end{equation}
(Lee e Weinberg 1977 ).\\
There are confirmed experimental evidence of the existence of massive neutrinos. In 1980, Lyubimov et al announced the detection of an electronic antineutrino with mass 30 eV, by means of the shape of the electron energy spectrum in the $\beta$ decay of tritium (N.A.Jelley 1986 ).
Experiments (Super-Kamiokande, SNOW) have obtained some evidence 
of non-zero mass from neutrino oscillations. This yields a difference of square 
masses of order $10^{-3}$ eV, and a mass of 0.05 eV (in the simplest case) (see Zuber 1998). \\
Among typical examples of CDM we have WIMPS and in particular axions and neutralinos (SUSY particle). This particle was postulated in order to solve the strong CP problem in nuclear physics. This problem arises from the fact that some interactions violate the parity, P, time inversion, T, and CP. If these are not eliminated, they give rise to a dipole momentum for the neutron which is in excess of ten order of magnitude with respect to experimental limits (Kolb e Turner 1990). The solution to the problem was proposed by Peccei-Quinn in 1977 in terms of a spontaneous symmetry breaking scheme. To this symmetry breaking should be associated (as shown by Weiberg \& Wilczec 1978) a Nambu-Goldstone boson: the axion. The axion mass ranges between $ 10^{-12} ev $-$ 1 Gev $. In cosmology there are two ranges of interest: $ 10^{-6} ev \leq m_{a} \leq 10^{-3}ev $ ; $ 3 ev \leq m_{a} \leq 8 ev $.
Axion production in the quoted range can originate due to a series of astrophysical processes (Kolb \& Turner 1990) and several are the ways these particles can be detected. Nevertheless the effort of researchers expecially in USA, Japan and Italy, axions remain hypothetical particles. They are in any case the most important CDM candidates.\\

In the following, I am going to speak about the basic ideas of structure formation. I shall write about density perturbations, their spectrum and evolution, about correlation functions and their time evolution, etc.\\

{\bf 1.3 Origin of structures.\\}

Observing our universe, we notice a clear evidence of inhomogeneity when we consider small scales (Mpcs). In clusters density reaches values of $10^3$ times larger than the average density, and in galaxies it has values $10^5$ larger than the average density (Kolb \& Turner 1990). If we consider scales larger than $10^2$ Mpcs universe appears isoptric as it is observed in the radio-galaxies counts, in CMBR, in the X background (Peebles 1971). The isotropy at the decoupling time, $t_{dec}$, at which matter and radiation decoupled, universe was very homogeneous, as showed by the simple relation:
\begin{equation}
\frac{\delta \rho}{\rho} = const \frac{\delta T}{T}
\end{equation}
(Kolb e Turner 1990) \footnote{In fact, COBE data gives $\frac{\delta T}{T} \leq 10^{-5}$}. The difference between the actual universe and that at decoupling is evident. The transformation between a highly homogeneous universe, at early times, to an highly local non homogeneous one, can be explained supposing that at $t_{dec}$ were present small inhomogeneities which grow up because of the gravitational instability mechanism (Jeans, J.H.,1902). Events leading to structure formation can be enumerated as follows:\\
(a) Origin of quantum fluctuations at Planck epoch.\\
(b) Fluctuations enter the horizon and they grow linearly till recombination.\\
(c) Perturbations grow up in a different way for HDM and CDM in the post-recombination phase, till they reach the non-linear phase.\\
(d) Collapse and structure formation.\\ 
Before $ t_{dec} $ inhomogeneities in baryonic components could not grow because photons and baryons were strictly coupled. 
This problem was not present for the CDM component. Then CDM perturbations started to grow up before those in the baryonic component when universe was matter dominated.
The epoch $ t_{eq} \approx 4.4*10^{10}(\Omega_{0} h^{2})^{-2} sec $, at which matter and radiation density are almost equal, can be considered as the epoch at which structures started to form. The study of structure formation is fundamentally an initial value problem. Data necessary for starting this study are:\\
1) Value of $ \Omega_{0} $. In CDM models the value chosen for this parameter is 1, in conformity with inflationary theory predictions.\\
2) The values of $\Omega_i$ for the different components in the universe. For example in the case of baryons, nucleosynthesis gives us the limit $ 0.014 \leq \Omega_{b} \leq 0.15 $ while $ \Omega_{WIMPS} \approx 0.9 $. \\
3) The perturbation spectrum and the nature of perturbations (adiabatic or isocurvature). The spectrum generally used is that of Harrison-Zeldovich: $ P(k) = Ak^{n} $ with $ n=1 $. The perturbation more used are adiabatic or curvature. This choice is  dictated from the comparison between theory and observations of CMBR anisotropy.\\

{\bf 1.4 The spectrum of density perturbation. \\}

In order to study the distribution of matter density in the universe it is generally assumed that this distribution is given by the superposition of plane waves independently evolving, at least until they are in the linear regime (this means till the overdensity $ \delta = \frac{\rho -\overline{\rho} }{\overline{\rho}}<<1 $). 
Let we divide universe in cells of volume $V_u$ and let we impose periodic conditions on the surfaces. If we indicate with 
$ \overline{\rho} $ the average density in the volume and with $ \rho({\bf r})$ the density in $ {\bf r} $, it is possible to define the density contrast as: 
\begin{equation}
\delta( {\bf r} ) = \frac{\rho( {\bf r}) - \overline{\rho}}{\overline{\rho}}
\end{equation}
This quantity can be developed in Fourier series:
\begin{equation}
\delta({\bf r}) = \sum_{{\bf k}} \delta_{{\bf k}}
exp(i{\bf k} {\bf r}) = \sum_{{\bf k}} \delta_{{\bf k}}
exp(-i{\bf k} {\bf r} ) \label{eq:sovra}
\end{equation}
( Kolb e Turner 1990), where $ k_{x} = \frac{2 \pi n_{x}}{l} $ (and similar conditions for the other components) and for the periodicity condition $ \delta(x,y,L) = \delta(x,y,0)$ (and similar conditions for the other components). 
Fourier coefficients $ \delta_{{\bf k}}$ are complex quantities given by: 
\begin{equation}
\delta_{{\bf k}} = \frac{1}{V_{u}} \int_{V_{u}} \delta({\bf r} )
exp(-i{\bf k}{\bf r}) d {\bf r}
\end{equation}
For mass conservation in $V_u$ we have also $ \delta_{{\bf k}=0}=0 $ while for reality of  $ \delta({\bf r}) $,  $ \delta_{{\bf k}}^{\ast} = \delta_{-{\bf k}} $.
If we consider n volumes, $V_u$, we have the problem of determining the distribution of Fourier coefficients $ \delta_{{\bf k}} $ and that of  $ \left|\delta\right| $.
We know that the coefficients are complex quantities and then $ \delta_{{\bf k}} = \left|\delta_{{\bf k}}
\right| exp (i \theta_{{\bf k}})$. If we suppose that phases are random, in the limit $ V_{u} \rightarrow \infty $ 
it is possible to show that  
we get 
$ \left|\delta\right|^{2} =
\sum_{{\bf k}} \left|\delta_{{\bf k}}\right|^{2} $. The Central limit theorem leads us to conclude that the distribution for 
$ \delta $ is Gaussian: 
\begin{equation}
P ( \delta ) \propto exp(\frac{-\delta^{2}}{\sigma^{2}} ) \label{eq:gau}
\end{equation}
(Efstathiou 1990).
The quantity $\sigma$ that is present in Eq. (\ref{eq:gau}) is the variance of the density field and is defined as: 
\begin{equation}
\sigma^{2} = < \delta ^{2} > = \sum_{{\bf k}} <
\left|\delta_{{\bf k}}\right|^{2} > = \frac{1}{V_{u}}
\sum_{{\bf k}}  \delta_{k}^{2}
\end{equation}
This quantity characterizes the amplitude of the inhomogeneity of the density field. If $ V_{u} \rightarrow \infty$, we obtain the more usual relation:
\begin{equation}
\sigma^{2} = \frac{1}{\left(2 \pi\right)^3} \int P(k) d^{3} k =
\frac{1}{2 \pi^{2}} \int P(k) k^{2} d k
\end{equation}
The term $ P( k ) = < \left|\delta\right|^{2}> $ is called ``Spectrum of perturbations". It is function only of k because the ensemble average in an isotropic universe depends only on r. A choice often made for the primordial spectrum is $ P( k) = A k^{n} $ which in the case $ n = 1 $ gives the scale invariant spectrum of Harrison-Zeldovich. An important quantity connected with the spectrum is the two-points correlation function $ \xi({\bf r}, t) $. It can be defined as the joint probability of finding an overdensity $\delta$ in two distinct points of space:
\begin{equation}
\xi({\bf r}, t) = < \delta( {\bf r},t )
\delta({\bf r}+{\bf x}, t) > \label{eq:corr}
\end{equation}
(Peebles 1980), where averages are averages on an ensemble obtained from several realizations of universe. Correlation function can be expressed as the joint probability of finding a galaxy in a volume $ \delta V_{1} $ and another in a volume 
$ \delta V_{2} $ separated by a distance $ r_{12} $:
\begin{equation}
\delta ^{2} P = n_{V}^{2} [1+ \xi(r_{12} )] \delta V_{1} \delta V_{2}
\end{equation}
where $ n_{V} $ is the average number of galaxies per unit volume. The concept of correlation function, given in this terms, can be enlarged to the case of three or more points.\\
Correlation functions have a fundamental role in the study of clustering of matter. If we want to use this function for a complete description of clustering, one needs to know the correlation functions of order larger than two (Fry 1982). By means of correlation functions it is possible to study the evolution of clustering. The correlation functions are, in fact, connected one another by means of an infinite system of equations obtained from moments of Boltzmann equation which constitutes the BBGKY (Bogolyubov-Green-Kirkwood-Yvon) hierarchy (Davis e Peebles 1978). 
This hierarchy can be transformed into a closed system of equation using closure conditions. Solving the system one gets information on correlation functions.\\
In order to show the relation between perturbation spectrum and two-points correlation function, we introduce in 
Eq. (\ref{eq:corr}), Eq. (\ref{eq:sovra}), recalling that $ \delta_{{\bf k}}^{\ast} = \delta( -{\bf k} )$ and taking the limit $ V_{u} \rightarrow \infty$, the average in the Eq. (\ref{eq:corr}) can be expressed in terms of the integral: 
\begin{equation}
\xi ( {\bf r} ) = \frac{1}{( 2 \pi ) ^{3}}
\int |\delta( {\bf k} )|^{2} exp( -i {\bf k} {\bf r} ) d^{3} k
\end{equation}
This result shows that the two-point correlation function is the Fourier transform of the spectrum. In an isotropic universe, it is $ |{\bf r}| = r $ and then $ | {\bf k} | =k $ and the spectrum can be obtained from an integral on $ |{\bf k}| = k $. Then correlation function may be written as:
\begin{equation}
\xi( r ) = \frac{1}{2 \pi^{2} }
\int k^{ 2} P (k ) \frac{sin( k r )}{k r} d k
\end{equation}
During the evolution of the universe and after perturbations enter the horizon, the spectrum is subject to modulations because of physical processes characteristic of the model itself (Silk damping (Silk 1968) for acollisional components, free streaming for collisional particles, etc.). These effects are taken into account by means of the transfer function
$ T(k;t) $ which connects the primordial spectrum $ P( k; t_{p} )$ at time $t_p$ to the final time $t_f$:
\begin{equation}
P(k;t_{f}) = \left[\frac{b(t_{f})}{b(t_{p})}\right]^{2}
T^{2}(k;t_{f}) P( k;t_{p})
\end{equation}
where b(t) is the law of grow of perturbations, in the linear regime. In the case of CDM models the transfer function is:
\begin{equation}
 T( k ) = \left\{1+\left[ak+(bk)^{1.5}+(ck)^{2}
\right]^{\nu}\right\}^{\frac{-1}{\nu}}
\end{equation}
(Bond e Efstathiou 1984), where $a=6.4 (\Omega h^{2} )^{-1}Mpc $; $b=3.0 (\Omega h^{2} )^{-1} Mpc$;
$ c=1.7 (\Omega h^{2} )^{-1} Mpc $; $ \nu=1.13$.
It is interesting to note that Eq. (\ref{eq:gau}) is valid only if $ \sigma << 1 $, since $ \left|\delta\right| \leq 1 $. This implies than non-linear perturbations, $ \sigma >> 1 $, must be non-Gaussian.
In fact when the amplitudes of fluctuations grow up, at a certain point modes are no longer independent and start to couple giving rise to non-linear effects that change the spectrum and correlation function (Juskiweicz et al 1984 ).
There are also some theories (e.g., cosmic strings (Kibble \& Turok 1986)) in which even in the linear regime perturbations are not Gaussian. 

{\bf 1.5 Curvature and isocurvature perturbations.\\}

The study of the evolution of density perturbations can be divided into two phases:\\
1) perturbations are outside the horizon, in other terms they have a scale $ \lambda $ larger than Hubble radius $ r_{H} = ct $ or $ \lambda \geq H^{-1} $;\\
2) perturbations are inside the horizon, $\lambda \leq H^{-1} $.\\
In studying the first case it is necessary to use general relativity and one can demonstrate that two different kinds of fluctuations exist: curvature or adiabatic (motivated by the simplest 
models of inflation) and isocurvature or isothermal. Curvature fluctuations are characterized by a fluctuation in energy density or in space curvature. For them, we can write:
\begin{equation}
 \frac{\delta S}{S} = \frac{3}{4} \frac{ \delta \rho_{r}}{\rho_{r}}
-\frac{\delta \rho_{m}}{\delta_{m}} = 3 \frac{\delta T}{T}-
\frac{\delta \rho_{m}}{\rho_{m}} = 0
\end{equation}
(Efstathiou 1990), where with m has been indicated the matter component, with r radiation, with S entropy and with T temperature. Last equation explains why these fluctuations are named adiabatic, since for them the entropy variation is zero.\\
Isocurvature or isothermal perturbations are not characterized by fluctuations in the curvature of the metric, but they are fluctuations in the local equation of state of universe and in agreement with the name it results $ \delta T = 0$. 
Until fluctuations of isocurvature are not inside the horizon, the causality principle does not permit a redistribution of energy density. This is possible only when perturbations enter the Horizon and isocurvature perturbations can be converted into perturbations in the energy density. As a consequence, the distinction between those two kinds of perturbations is no longer meaningful after them enter the horizon (Suto et al 1985; Gouda \& Sasaki 1986).
The origin of curvature perturbations may be explained inside inflationary models or assuming that they are initially present as perturbations of the metric. Isocurvature fluctuations may be always produced in inflationary scenarios from fluctuations in the density number of barions or axions (Efstathiou 1990, Kolb e Turner 1990 ).\\

{\bf 1.6 Perturbations evolution.\\}

Density perturbations in the components of the universe evolve with time. In order to get the evolution equations for $\delta$ in Newtonian regime, it is possible to use several models. In our model, we assume that gravitation dominates on the other interactions and that particles (representing galaxies, etc.) move collisionless in the potential $\phi$ of a smooth density function (Peebles 1980). The distribution function of particles for position and momentum is given by:
\begin{equation}
d N = f( {\bf x}, {\bf p},t) d^{3}x d^{3} p
\end{equation}
and density:
\begin{equation}
\rho({\bf x} , t )= m a^{-3} \int d^{3} p f({\bf x}, {\bf p}, t)=
\rho_{b}\left[1+ \delta({\bf x}, t)\right] \label{eq:densi}
\end{equation}
where m is the mass of a particle and $\rho_{b}$ the background density. Applying Liouville theorem to the probability density on a limited region of phase-space of the system we have that f verifies the equation:
\begin{equation}
\frac{\partial f}{\partial t }+\frac{{\bf p}}{m a^{2}}
\bigtriangledown f - m \bigtriangledown \phi \frac{
\partial f}{\partial {\bf p}} = 0 \label{eq:liou}
\end{equation}
The distribution function f that appears in the previous equations cannot be obtained from observations. It is possible to measure moments of f (density, average velocity, etc.). We want now to obtain the evolution equations for $\delta$. For this reason, we start integrating Eq. (\ref{eq:liou}) on $ {\bf p} $ and after using Eq. (\ref{eq:densi}), we get:
\begin{equation}
a^{3} \rho_{b} \frac{\partial \delta}{\partial t}+
\frac{1}{a^{2}} \bigtriangledown \int {\bf p} f d^{3} p =0 \label{eq:tre}
\end{equation}
If we define velocity as: 
\begin{equation}
{\bf v} = \frac{\int \frac{{\bf p}}{m a}f d^{3} p}{\int f d^{3} p}
\end{equation}
and introduce it in Eq. (\ref{eq:tre}) we get:
\begin{equation}
\rho_{b} \frac{\partial \delta}{\partial t}+
\frac{1}{a} \bigtriangledown (\rho {\bf v} )= 0
\end{equation}
We can now multiply Eq. (\ref{eq:liou}) for $ {\bf p} $ and integrate it on the momentum:
\begin{equation}
\frac{\partial}{\partial t} \int p_{\alpha} f d^{3} p +
\frac{1}{m a^{2}} \partial_{\beta} \int p_{\alpha}
p_{\beta} f d^{3} p + a^{3} \rho ( {\bf x}, t) \phi_{, \alpha} =0
\end{equation}
this last in Eq. (\ref{eq:tre}) leaves us with:
\begin{equation}
\frac{\partial^{2} \delta }{\partial t^{2}} + 2\frac{\dot{a}}{a}
\frac{\partial \delta}{\partial t} = \frac{1}{a^{2}}
\bigtriangledown \left[(1+\delta ) \bigtriangledown \phi\right]+
\frac{1}{\rho_{b} m a^{7}} \partial_{\alpha} \partial_{\beta}
\int p_{\alpha} p_{\beta} \phi d^{3} p
\end{equation}
and finally using 
\begin{equation}
< v_{\alpha} v_{\beta} > = \frac{\int f p_{\alpha} p_{\beta} d^{3} p }
{m a^{2} \int f d^{3} p}
\end{equation}
the equation for the evolution of overdensity becomes:
\begin{equation}
\frac{\partial^{2} \delta }{\partial t^{2}} + 2\frac{\dot{a}}{a}
\frac{\partial \delta}{\partial t} = \frac{1}{a^{2}}
\bigtriangledown \left[(1+\delta ) \bigtriangledown \phi\right]+
\frac{1}{a^{2}} \partial_{\alpha} \partial_{\beta}
\left[(1+\delta) < v^{\alpha} v^{\beta} >\right]
\end{equation}
(Peebles 1980). The term $ < v_{\alpha} v_{\beta} > $ is the tensor of anisotropy of peculiar velocity. This is present in the gradient and then it behaves as a pressure force. If we consider an isolated and spherical perturbation, it is possible to assume that initial asymmetries does not grow up and so we can suppose, in this hypothesis that $ < v_{\alpha} v_{\beta} > = 0 $. In this case and with the linearity assumption $ \delta << 1 $ we have:
\begin{equation}
\frac{\partial^{2} \delta}{\partial t^{2}} + 2\frac{\dot{a}}{a}
\frac{\partial \delta}{\partial t } = 4 \pi G \rho_{b} \delta
\end{equation}
This equation in an Einstein-de Sitter universe ($\Omega = 1$,
$\Lambda = 0 $) has the solutions:
\begin{equation}
\delta_{+}= A_{+}({\bf x}) t^{\frac{2}{3}} \hspace{1cm}
\delta_{-} ({\bf x},t) = A_{-} ({\bf x}) t^{-1}
\end{equation}
The perturbation is then done of two parts: a growing one, becoming more and more important with time, and a decaying one
becoming negligible with increasing time, with respect to the growing one. 

In the case of open models with no cosmological constant: $\Omega<1$,
$\Lambda=0$, we can write:

\begin{equation} 
\frac{\dot{a}^2}{a^2}=\frac{8}{3}\pi G\bar{\rho}\left(1+
\left(\Omega_0^{-1} -1\right) a\right), \label{eq:frw_open} 
\end{equation}
\noindent 
and the $a(t)$ evolution can be expressed through the following
parametric representation:

\begin{eqnarray} 
a(\eta)&=&\frac{\Omega_0}{2(1-\Omega_0)}({\rm cosh} \eta -1) \\
     t(\eta)&=&\frac{\Omega_0}{2H_0(1-\Omega_0)^{3/2}}({\rm sinh}\eta -\eta).
\nonumber \label{eq:a_ev_open} 
\end{eqnarray} 
In the case of flat models with positive cosmological constant: $\Omega<1$,
$\Lambda\neq 0$, $\Omega+\Lambda/3H_0^2 = 1$, we can write:
\begin{equation} 
\frac{\dot{a}^2}{a^2}=\frac{8}{3}\pi G\bar{\rho}\left(1+
\left(\Omega_0^{-1} -1\right) a^3\right), \label{eq:frw_cosm} 
\end{equation} 
\begin{equation} 
a(t) = \left(\Omega_0^{-1}-1\right)^{-1/3}{\rm sinh}^{2/3}
\left(\frac{3}{2}\sqrt{\frac{\Lambda}{3}} t\right). \label{eq:a_ev_cosmo} 
\end{equation} 

Before concluding this section, we want to find an expression for the velocity field in the linear regime. 
Using the equation of motion $ {\bf p} = m a^{2} {\bf \dot{x}} $, $ \frac{d {\bf p}}{d t} = -m \bigtriangledown \phi $ and the proper velocity of a particle, $ v = a \bf{\dot{x}} $, verify the equation:
\begin{equation}
\frac{d {\bf v}}{d t} + {\bf v} \frac{\dot{a}}{a} =
-\frac{\bigtriangledown \phi}{a} = G \rho_{b} a
\int d^{3} x \delta({\bf x}, t) \frac{{\bf x} -{\bf x'}}{\left|{\bf x}-
{\bf x'}\right|}
\end{equation}
Supposing that ${\bf v} $ is a similar solution for the density, $ {\bf v} = {\bf V_{+}({\bf x}, t)} t^{p}$, we get:
\begin{equation}
v^{\alpha} = \frac{H a}{4 \pi} \frac{\partial}{\partial x^{\alpha}}
\int d^{3} x' \frac{\delta ( {\bf x'}, t)}{\left|{\bf x'}-
{\bf x}\right|}
\end{equation}
(Peebles 1980). 
This solution is valid just as that for $ \delta $ in the linear regime. At time $ t = t_{0} $ this regime is valid on scales larger than $ 8 h^{-1} Mpc $.\\

\begin{flushleft}
{\bf 1.7 Non-linear phase.\\}
\end{flushleft}
\vspace*{5.0mm}
Linear evolution is valid only if $ \delta << 1 $ or similarly, if the mass variance, $\sigma$, is much less than    
unity. When this condition is no longer verified (e.g., if we consider scales smaller than $8h^{-1}$ Mpc), it is necessary to develope a non-linear theory. In regions smaller than $8h^{-1}$ Mpc
galaxies are not a Poisson distribution but they tend to cluster.   
If one wants to study the properties of galactic structures or clusters of galaxies, it is necessary to introduce a non-linear theory of clustering. A theory of this last item is too complicated to be developed in a purely theoretical fashion. The problem can be faced assuming certain approximations that simplifies it (Zel'dovich 1970) or as often it is done, by using N-Body simulations of the interesting system.  
The approximations are often used to furnish the initial data to simulations. 
In the simulations, a large number of particles are randomly distributed in a sphere, in the points of a cubic grid, in order to eliminate small scale noise. The initial spectrum is obtained perturbing the initial positions by means of a superposition of plane waves having random distributed phases and wave vector (West et al. 1987).
Obviously, the universe is considered in expansion (or comoving coordinates are used), and then the equation of motion of particles are numerically solved. For what concerns the analytical approximations one of the most used is that of Ze`ldovich (1970). This gives a solution to the problem of the grow of perturbations in an universe with $p=0$ not only in the linear regime but even in the mildly non-linear regime.
In this approximation, one supposes to have particles with initial position given in Lagrangian coordinates $ {\bf q} $.
The positions of particles, at a given time t, are given by: 
\begin{equation}
{\bf x}={\bf q}+b(t){\bf p(q)}
\end{equation}
where ${\bf x}$ indicates the Eulerian coordinates, ${\bf p(q) } $ describes the initial density fluctuations and $b(t)$ describes their grow in the linear phase and it satisfies the equation:
\begin{equation}
\frac{d^{2}b}{dt^{2}}+2a^{-1}\frac{db}{dt}\frac{da}{dt}=4 \pi G
\rho b
\end{equation}
The equation of motion of particles, according to the quoted approximation, is given by: 
\begin{equation}
{\bf v} = \dot{a} {\bf q} + \dot{b} {\bf p}({\bf q} )
\end{equation}
The peculiar velocity of particles is given by:
\begin{equation}
{ \bf u}=\frac{d{\bf x}}{dt} =\frac{db}{dt}{ \bf p(q)}
\end{equation}
while the density of the perturbed system is given by:
\begin{equation}
\rho({\bf q},t)=\overline{\rho}
\left|\frac{\partial q_{j}}{\partial x_{k}}\right| =
\overline {\rho} \left| \delta_{jk} + b(t)
\frac{\partial p_{k} }{\partial q_{j} }\right|^{-1} \label{eq:cas}
\end{equation}
Developing the Jacobian present in Eq. (\ref{eq:cas}) at first order in $ b(t) {\bf p(q)} $, one obtains:
\begin{equation}
\frac{\delta \rho}{\overline{\rho}}\approx -b(t)
\bigtriangledown_{{ \bf q}}{\bf p(q)}
\end{equation}
This equation can be re-written, separating the space and time dependence, as in the equation for $ { \bf u} $, and writing:
\begin{equation}
b(t)=t^{\frac{2}{3}} \hspace{1.0cm} {\bf p(q)}=
\sum_{{\bf k}}i\frac{{\bf k}}{\left|{\bf k}\right|^{2}}A_{{\bf k}}
exp(i{\bf k q})
\end{equation}
in the form: 
\begin{equation}
\frac{\delta \rho}{\overline{\rho}}=
\sum_{{ \bf k}} A_{{\bf k}}t^\frac{2}{3} exp(i{\bf k q})
\end{equation}
(Efstathiou 1990), that leads us back to the linear theory. 
In other words, Ze`ldovich approximation is able to reproduce the linear theory, and is also able to give a good approximation in regions with $ \frac{\delta \rho}{\overline{\rho}}>>1$. Using the expression for $ p(q) $, the Jacobian in Eq. (\ref{eq:cas}) is a real matrix and symmetric that can be diagonalized. 
With this $ p(q) $ the perturbed density can be written as:
\begin{equation}
\rho({\bf q},t)=\frac{\overline{\rho}}{(1-b(t)\lambda_{1}(q))
(1-b(t)\lambda_{2}(q))(1-b(t)\lambda_{3}(q))} \label{eq:panc}
\end{equation}
where $ \lambda_{1} $, $ \lambda_{2} $, $ \lambda_{3} $ are the three eigenvalues of the Jacobian, describing the expansion and contraction of mass along the principal axes. From the structure of the last equation, we notice that in regions of high density Eq. (\ref{eq:panc}) becomes infinite and the structure of collapse in a pancake, in a filamentary structure or in a node, according to values of eigenvalues. Some N-body simulations (Efstathiou e Silk, 1983) tried to verify the prediction of 
Ze`ldovich approximation, using initial conditions generated using a spectrum with a cut-off at low frequencies. 
The results showed a good agreement between theory and simulations, for the initial phases of the evolution 
($ a(t)=3.6 $). Going on, the approximation is no more valid starting from the time of shell-crossing. After shell-crossing, particles does not oscillate any longer around the structure but they pass through it making it vanish. 
This problem has been partly solved supposing that particles, before reaching the singularity they sticks the one on the other, due to a dissipative term that simulates gravity and then collects on the forming structure.  
This model is known as ``adesion-model" (Gurbatov et al. 1985). \\
Summarizing, Zel'dovich approximation gives a description of the transition between linear and non-linear phase. It is expecially used to get the initial conditions for N-body simulations. \\

\begin{flushleft}
{\bf 1.8 Quasi-linear regime.\\}
\end{flushleft}

We have seen in the previous section that in the case of regions of dimension smaller than $ 8 h^{-1} Mpc $, the linear theory is no more a good approximation and  a new theory is needed or  N-body simulations. 
Non-linear theory is able to calculate quantities as the formation redshift of a given class of objects as galaxies and clusters, the number of bound objects having masses larger than a given one, the average virial velocity and the correlation function. It is possible to get an estimate of the given quantities as that of other not cited, using an intermediate theory between the linear and non-linear theory: the quasi-linear theory. This last is obtained adding to the linear theory a model of gravitational collapse, just as the spherical collapse model. Important results that the theory gives is the bottom-up formation of structures (in the CDM model). Other important results are obtained if we identify density peaks in linear regime with sites of structure formation. Two important papers in the development of this theory are Press-Schechter (1974) and that of Bardeen et al. (1986). This last paper is an application of the ideas of the quasi-linear theory to the CDM model. The principles of this approach are the following:\\
\begin{itemize}
\item
Regions of mass larger than M that collapsed can be identified with regions where the density contrast evolved according to linear regime, $ \delta ( M, x ) $, has a value larger than   a threshold, $ \delta_{c} $.
\item After collapse regions does not fragment.
\end{itemize}
The major drawbacks of the theory, as described in Bardeen et al. (1986) are fundamentally the fact that the estimates that can be obtained by means of this theory depends on the threshold $ \delta_{c} $, on the ratio between the filtering mass and that of objects and from other parameters. Nevertheless, this theory has helped cosmologists in obtaining estimate of important quantities as those previously quoted, and at same time give evidences that leads to exclude very low values for spectrum normalization.

{\bf 1.9 Spherical Collapse}

Spherical symmetry is one of the few cases in which gravitational
collapse can be solved exactly (Gunn \& Gott 1972; Peebles 1980).  In
fact, as a consequence of Birkhoff's theorem, a spherical perturbation
evolves as a FRW Universe with density equal to the mean density
inside the perturbation.

The simplest spherical perturbation is the top-hat one, i.e. a
constant overdensity $\delta$ inside a sphere of radius $R$; to avoid
a feedback reaction on the background model, the overdensity has to be
surrounded by a spherical underdense shell, such to make the total
perturbation vanish. The evolution of the radius of the perturbation
is then given by a Friedmann equation.

The evolution of a spherical perturbation depends only on its initial
overdensity. In an Einstein-de Sitter background, any spherical 
overdensity
reaches a singularity (collapse) at a final time:

\begin{equation} t_c=\frac{3\pi}{2}\left(\frac{5}{3}\delta(t_i)\right)^{-3/2} t_i.
\label{eq:spherical_coll} \end{equation}

\noindent
By that time its linear density contrast reaches the value: 

\begin{equation} \delta_l(t_c)=\delta_c=\frac{3}{5}\left(\frac{3\pi}{2}\right)^{3/2}
\simeq 1.69. \label{eq:delta_c_sph}\end{equation}

\noindent
In an open Universe not any overdensity is going to collapse: the
initial density contrast has to be such that the total density inside
the perturbation overcomes the critical density. This can be
quantified (not exactly but very accurately) as follows: the growing
mode saturates at $b(t)=5/2(\Omega_0^{-1}-1)$, so that a perturbation
ought to satisfy $\delta_l>1.69\cdot 2(\Omega_0^{-1}-1)/5$ to be able
collapse.

Of course, collapse to a singularity is not what really happens in
reality. It is typically supposed that the structure reaches virial
equilibrium at that time. In this case, arguments based on the virial
theorem and on energy conservation show that the structure reaches a
radius equal to half its maximum expansion radius, and a density
contrast of about 178. In the subsequent evolution the radius and the
physical density of the virialized structure remains constant, and its
density contrast grows with time, as the background density decays.
Similarly, structures which collapse before are denser than the ones
which collapse later.

Spherical collapse is not a realistic description of the formation of
real structures; however, it has been shown (see Bernardeau 1994 for
a rigorous proof or Valageas 2002a,b) that high peaks ($> 2\sigma$) follow spherical
collapse, at least in the first phases of their evolution. However, 
a small systematic departure from
spherical collapse can change the statistical properties of collapse
times.

Spherical collapse can describe the evolution of underdensities. A
spherical underdensity is not able to collapse (unless the Universe is
closed!), but behaves as an open Universe, always expanding unless its
borders collide with neighboring regions. At variance with
overdensities, underdensities tend to be more spherical as they
evolve, so that the spherical model provides a very good approximation
for their evolution.

\begin{flushleft}
{ \bf 1.10 Mass function.\\}
\end{flushleft}

One of the most important quantities in cosmology is the mass function or multiplicity function. It can be described by the relation:
\begin{equation}
dN = n( M ) dM
\end{equation}
that is the number of objects per unit volume, having a mass in the range $ M $ ed $ M + dM $. The multiplicity function can also be used to define the luminosity function after having fixed the ratio $ \frac{ M}{ L} $.
Obtaining the mass function starting from that of luminosity is complicated since the ratio $ \frac{ M }{ L } $
is known with noteworthy uncertainty and it is different for different objects and moreover the luminosity function for objects like galaxies depend on the morphological type  (Binggeli \& Tamman 1985). Finally trying to determine the luminosity function observatively is problematic (see for example G.Efstathiou, R.S.Ellis, B.A.Peterson (1988)).\\
For the above reasons, the theoretical determination of the mass function is very important. One of the most successful study in the subject is that of Press \& Schechter (1974). This theory is based upon these hypotheses:
\begin{itemize}
\item
The linear density field is described by a stochastic Gaussian field. The statistics of the matter distribution is Gaussian.
\item 
The evolution of density perturbations is that described by the linear theory. Structures form in those regions where the overdensity linearly evolved and filtered with a top-hat filter exceeds a threshold $ \delta_{c} $ ($ \delta_{c} =1.68 $, obtained from the spherical collapse model (Gunn \& Gott 1972)). 
\item 
for $ \delta \geq \delta_{c} $ regions collapse to points. The probability that an object forms at a certain point is proportional to the probability that the point is in a region with $ \delta \geq \delta_{c} $ given by:
\begin{equation}
P ( \delta , \delta_{c} )=
\int_{\delta_{c}}^{\infty}d \delta \frac{1}{\sigma
(2 \pi)^{\frac{1}{2}}}
 exp \left(-\frac{\delta^{2}}{2 \sigma^{2}}\right)
\end{equation}
The multiplicity function is given by:
\begin{equation}
\rho ( M, z )= - \rho_{0}\frac{ \partial P}{\partial M } dM =
n( M ) M dM \label{eq:ps}
\end{equation}
\end{itemize}
If we add the conditions
$ \Omega=1 $, $ \left|\delta_{k}\right|^{2} \propto k^{n}$,
the Press-Schechter solution is autosimilar and has the form:
\begin{eqnarray}
\rho ( M, z) =\frac{ \rho}{\sqrt{2 \pi}}
\left(\frac{n+3}{3}\right) \left( \frac{M}{M_{\ast }}( z)
\right)^{\frac{n+3}{6}} \nonumber \\
\times exp \left[ -\frac{1}{2} \frac{M}{M_{\ast}}( z )^{
\frac{n+3}{3}}\right]\frac{dM }{M}
\end{eqnarray}
where $ M_{\ast}(z ) \propto (1+z)^{-\frac{6}{n+3}} $. 
Several are the problems of the theory: 
\begin{itemize}
\item 
{\bf Statistical problems:} in the limit of vanishing smoothing radii,
or of infinite variance, the fraction of collapsed mass, 
asymptotes to 1/2. This is a signature of
linear theory: only initially overdense regions, which constitute half
of the mass, are able to collapse. Nonetheless, underdense regions can
be included in larger overdense ones, or, more generally,
non-collapsed regions have a finite probability of being included in
larger collapsed ones; this is commonly called {\it cloud-in-cloud
problem}.  PS argued that the missing mass would accrete on the formed
structures, doubling their mass without changing the shape of the MF;
however, they did not give a true demonstration of that. Then, they
multiplied their MF by a ``fudge factor'' 2. Other authors (see
Lucchin 1989) used to multiply the MF by a factor $(1+f)$, with $f$
denoting the fraction of mass accreted by the already formed
structures.
\item
{\bf Dynamical problems:} the heuristic derivation of the PS MF
bypasses all the complications related to the highly non-linear
dynamics of gravitational collapse. Spherical
collapse helps in determining the $\delta_c$ parameter and in identifying
collapsed structures with virialized halos. However, the PS procedure
completely ignores important dynamical elements, such as the role of
tides and the transient filamentary geometry of collapsed structures.
Moreover, supposing that every structure virializes just after
collapse is a crude simplification: when a region collapses, all its
substructure is supposed by PS to be erased at once, while in
realistic cases the erasure of substructures is connected to the
two-body interaction of already collapsed clumps, an important piece
of gravitational dynamics which is completely missed by the PS
procedure.
\item
{\bf Geometrical problems:} to estimate the mass function from
the fraction of collapsed mass at a given scale it is
necessary to relate the mass of the formed structure to the resolution

In practice,
the true geometry of the collapsed regions in the Lagrangian space
(i.e. as mapped in the initial configuration) can be quite complex,
especially at intermediate and small masses; in this case a different
and more sophisticate mass assignment ought to be developed, so that
geometry is taken into account. For instance, if structures are
supposed to form in the peaks of the initial field, a different and
more geometrical way to count collapsed structures could be based on
peak abundances.
\end{itemize}

Despite all of its problems, the PS procedure proved successful, as
compared to N-body simulations, and a good starting point for all the
subsequent works on the subject (Efstathiou et al. (1988); Efstathiou \& Rees (1988), 
Bond et al. (1991), 
White, Efstathiou \& Frenk (1993),
Jain \& Bertschinger (1994), 
Lacey \& Cole (1994), Efstathiou (1995), 
Bond \& Myers (1996b).  Most authors reported the PS
formula to fit well their N-body results; nonetheless, all the authors
agree in stating the validity of the PS formula to be only
statistical, i.e.  the existence of the single halos is not well
predicted by the linear overdensity criterion of PS (see in particular
Bond et al. 1991)). 
There are however some exceptions to this general agreement: Brainerd
{\&} Villumsen (1992) reported their MF, based on a CDM spectrum, to
be very similar to a power-law with slope $-2$, different from the PS
formula both at small and at large masses.  Jain \& Bertschinger
(1994), Gelb \& Bertschinger (1994) and Ma \& Bertschinger (1994)
noted that, to make the PS formula agree with their simulations (based
on CDM or CHDM spectra), it is necessary to lower the value of the
$\delta_c$ parameter as redshift increases.  The same thing was found by
Klypin et al. (1995), but was interpreted as an artifact of their
clump-finding algorithm. Recent simulations seem to confirm this trend.

Lacey \& Cole (1994) extended the comparison to N-body simulations to
the predictions for merging histories of dark-matter halos; they found again a good agreement between theory and
simulations. This fact is noteworthy, as merging histories contain
much more detailed information about hierarchical collapse.
Several improvements of the theory exists: Bond et al. 1991; Lacey \& Cole (1993) (the so called Extended Press \& Schechter (EPS) formalism; Sheth \& Tormen (2002). 
 
{\bf 1.11 CDM and HDM cosmogony.\\}

The study of origin and formation of structure in the universe has been historically fundamentally framed into two theories: the CDM theory, in which WIMPS constitutes the main part of Dark Matter, and HDM in which neutrinos dominate. As we are going to see, structure formation in these scenarios is completely different since WIMPS and neutrinos are subject to different physical phenomena and then the transfer function is noteworthy different in the two cases. 
Both theories have the same starting points:\\
1) The universe is fundamentally constituted by Dark Matter (WIMPS in the CDM model and neutrinos
in the HDM) and $\rho =\rho_{c} $ (  $\Omega =1$). \\
2) Baryons give a small contribution to the mass of the universe.\\
3) Fluctuations originating in the primordial universe are adiabatic, scale invariant, $n=1$, and Gaussian (Efstathiou 1990).\\
If universe is dominated by neutrinos with mass $ m_{\nu} = 30 ev $ the transfer function is determined by the free-streaming (or Landau damping) of neutrinos. This phenomenon consists in the smoothing of inhomogeneities in the primordial universe (due to perturbations in the acollisional components) because of the motion of neutrinos from overdense to underdense regions. Neutrinos diffusion and the smoothing of inhomogeneities is possible only before $ t = t_{eq} $.
After this epoch, there is no longer free-streaming but the density perturbation has definitely changed by its previous action. Free-streaming scale or mass can be estimated calculating the distance covered from a particle decoupled from plasma. Results that one obtains for the free-streaming scale and mass is (Kolb \& Turner 1990): 
\begin{equation}
\lambda_{FS} \approx 40 Mpc (m_{\nu}/30 ev )^{-1} \hspace{1cm}
M_{FS} \approx 10^{15}(m_{\nu}/ 30 ev)^{-2} M_{0}
\end{equation}
Because of free-streaming of neutrinos the HDM spectrum is characterized by a cut-off at short wave-length. The result is that the first objects that form are superclusters and structure formation proceeds because of fragmentation. 
Zel'dovich (1970) showed that the first structure to form are flat and were called "pancakes". After formation, these objects enter in the non-linear phase along one of the axes and baryons inside start to collide and dissipate their gravitational energy. Galaxies form for fragmentation processes. Structure formation follows a 'top-down' scheme, that means that larger objects (e.g., clusters) form before, while smaller objects (e.g., galaxies) later. N-body simulations of HDM universes (Klypin e Shandarin 1983; Frenk et al 1983; Centrella and Mellot 1983) showed that on scales larger than 10 Mpc structures is qualitatively similar to voids and to the filamentary structure that is visible in the CFA, but the clustering measured in N-body simulations is larger than that observed in the CFA.
When one tries to reproduce the observed correlation function, one arrives to the conclusion that pancaking should have happened at redshift $ z \leq 1 $, in disagreement with observed galaxies having $ z \geq 1 $ and QSO with $ z \geq 3 $.
A further problem of the model is that of the peculiar velocity that are smaller than values obtained from observations (Kolb e Turner 1990). \\
The HDM model after a series of studies in the '80s has been abandoned for the problems it has and replaced by another model, the CDM, which is in better agreement with observations. The CDM model has a spectrum without a cut-off at short wave -length (at least till scales much smaller than galactic scales) because the damping scale is unimportant for WIMPS with mass $ > $ 1 Gev. Structure formation is typically hyerachical: from smaller scale structure to larger ones. This scheme is  a 'bottom-up' scheme. When the CDM model was introduced it obtained noteworthy successes in the description of the characteristic of the universe (clustering statistics of galaxies, peculiar velocities, CMBR fluctuations) from the galactic scale on ( Peebles 1982 b; Blumenthal et al 1984; Bardeen et al 1986;
White et al 1987; Frenk et al 1988; Efstathiou 1990). The model has shown some weak points, when compared with more and more precise data.\\
The reason of the success of the CDM model is fundamentally due to the fact that WIMPS interact with matter by means of gravity only, and does not feel the effect of pressure forces due to interaction with radiation (to which matter components are subject). Structure formation starts before in the CDM component, at $ t< t_{eq} $, which give rise to the potential wells in which baryonic matter can then fall. It is important to notice that, in order to reproduce observations, an additive hypothesis must be added: the biasing hypothesis, that can be summarized in: {\it light does not trace mass}. \footnote{If this hypothesis is not introduced the discrepancies between data and theory can be reduced assuming a value of $ H_{0} \approx 25 km/Mpc s $}. Typical problems of the CDM model in absence of biasing are the too high values for the r.m.s. of peculiar velocity of couple of galaxies (values of 1000 km/s vs. $ 300 \pm 50 $ km/s observed).
Another problem is the correlation length, $r_{0} $, in N-body simulations, for the correlation function $ \xi $ which is equal to $ 1.3 h^{-1} Mpc $, smaller than the observed value $ 5.5 h^{-1} Mpc $ ( Davis et al. 1985). 
On the other side, if, in order to eliminate the quoted problems one introduce the biasing hypothesis, there is the supplementary problem of finding a physical mechanism that explains the origin of bias. Several conjectures has been proposed (Rees 1985; Silk 1985; Dekel e Rees 1987) but there is no full agreement on them. 

Summarizing, one can tell that
although at the beginning the standard form of CDM was very successful in describing the structures observed in the
Universe (galaxy clustering statistics, structure formation epochs, peculiar velocity flows) (Peebles, 1982; Blumenthal et al. 1984; Bardeen et al. 1986 -hereafter BBKS; White et al. 1987; Frenk et al. 1988; Efstathiou 1990) recent measurements
have shown several deficiencies in the model, at least, when any bias of the distribution of galaxies relative to the mass is constant with scale (see Babul \& White 1991; Bower et al. 1993; Del Popolo \& Gambera 1998a,b). Some
of the most difficult problems that must be reconciled  with the theory are:
\begin{itemize}
\item the magnitude of the
dipole of the angular distribution of optically selected galaxies (Lahav et
al. 1988; Kaiser \& Lahav 1989);
\item the possible observations of clusters of
galaxies with high velocity dispersion at $z\geq 0.5$ (Evrard 1989);
\item the
strong clustering of rich clusters of galaxies, $ \xi_{cc}(r) \simeq (r / 25h^{-1}Mpc)^{-2}$, 
far in excess of CDM predictions (Bahcall \& Soneira 1983);
\item the X-ray temperature distribution function of clusters,
over-producing the observed clusters abundances (Bartlett \& Silk 1993); 
\item the
conflict between the normalization of the spectrum of the perturbation which is required by different types of observations;
\item 
the incorrect scale dependence of the galaxy correlation
function, $\xi (r)$, on scales $10$ to $100$ $h^{-1} Mpc$, having $\xi (r)$ too little power on the large scales compared to the power on smaller scales (Lahav et al. 1989; Maddox et al. 1990a; Saunders et al. 1991; 
Peacock 1991; Peacock \& Nicholson 1991).
\item
Normalization obtained from COBE data (Smoot et al. 1992) on scales of the order of $10^3Mpc$ 
requires $\sigma_8=0.95\pm 0.2$, where $\sigma_8$ is the rms value of $\frac{\delta M}{M}$ in a sphere of $8 h^{-1}$Mpc.
Normalization on scales $10$ to $50Mpc$ obtained from QDOT (Kaiser et al. 1991) and POTENT (Dekel et al. 1992) requires that $\sigma _8$ is in the range $0.7\div 1.1$, which is compatible with COBE normalization
while the observations of the pairwise velocity dispersion of galaxies on
scales $r\leq 3Mpc$ seem to require $\sigma _8<0.5$. 
\item 
Another problem of CDM
model is the incorrect scale dependence of the galaxy correlation
function, $\xi (r)$, on scales $10$ to $100$ $Mpc$, having $\xi (r)$ too
little power on the large scales compared to the power on smaller scales.
The APM survey (Maddox et al. 1990), giving the galaxy angular
correlation function, the 1.2 Jy IRAS power spectrum, the QDOT survey
(Saunders et al. 1991), X-ray observations (Lahav et al. 1989) and radio
observations (Peacock 1991; Peacock \& Nicholson 1991) agree with the quoted
conclusion. As shown in recent studies of galaxy clustering on large scales
(Maddox et al. 1990; Efstathiou et al. 1990b; Saunders et al. 1991) the
measured rms fluctuations within spheres of radius $20h^{-1}Mpc$ have
value 2-3 times larger than that predicted by the CDM model.\\
\item Density profiles of CDM halos: the cusp obtained from numerical 
simulations seems too steep. 
-\item simulations might yield too many satellites for galaxies like our own. 
Though this second problem may have been the result of bad comparison 
of simulations with observations. 
This yielded a surge of interest in the last 2-3 years for Warm Dark matter and 
slightly collisional matter.
\end{itemize}

These discrepancies between the theoretical predictions of the CDM model
and  the observations led many authors to conclude that the shape of the
CDM spectrum is incorrect and to search alternative models
(Peebles 1984; Shafi \& Stecker 1984; Valdarnini \& Bonometto 1985; 
Holtzman 1989; Efstathiou et al. 1990a; Turner 1991; Schaefer 1991; Cen et al. 1992; Schaefer \& Shafi 1993; Holtzman \& Primack 1993; Bower et al. 1993).\\
Alternative models with more large-scale power than CDM have been introduced
in order to solve the latter problem. Several authors (Peebles 1984;
Efstathiou et al. 1990a; Turner 1991) have
lowered the matter density under
the critical value ($ \Omega_m < 1$) and have added a
cosmological constant in order to retain 
a flat Universe ($ \Omega _m + \Omega _\Lambda = 1$) .
The spectrum of the matter density is specified by the
transfer function, but its shape is
affected because of the fact that the epoch of
matter-radiation equality is earlier,
$1+z_{eq}$ being increased by a factor $1/\Omega_{m}$. 
The epoch of
matter-radiation equality is earlier, because $1+z_{eq}$ is increased by a
factor $1/\Omega _m$.
Around the epoch $z_\Lambda $ the growth of the
density contrast slows down and ceases after $z_\Lambda $.
As a consequence the
normalization of the transfer function begins to fall, even if its shape
is retained (and pushes its imprint to larger scales). 
Mixed Dark Matter models (MDM) (
Bond et al. 1988; 
Shafi \& Stecker 1984; Valdarnini \& Bonometto 1985; 
Schaefer et al. 1989; 
Holtzman 1989;
Schaefer 1991; Shaefer \& Shafi 1993; Holtzman \& Primack 1993) increase
the large-scale power because neutrinos free-streaming damps the power on
small scales. Alternatively, changing the primeval spectrum
several problems of CDM are
solved (Cen et al. 1992). Finally it is possible to assume
that the threshold for galaxy formation
is not spatially invariant but weakly modulated ($2\%-3\%$ on scales $%
r>10h^{-1}Mpc$) by large scale density fluctuations, with 
the result that the clustering on
large-scale is significantly increased (Bower et al. 1993). \\

{\bf Conclusions.\\}

This paper provides a review of the variants of dark matter (CDM, HDM) which are thought to be fundamental components of the universe and their role in origin and evolution of structures. Studies of several decades have shown that, if we have a right knowledge of the law of gravity, dark matter is a fundamental component of our universe. While models based upon Hot Dark Matter (e.g., neutrinos) gives a reasonable description of structures on large scales
models based upon Cold Dark Matter (e.g., axions) are more successful 
in describing small and intermediate scales.
A fundamental ingredient in the recipe of structure formation is 
inflation which provides a spectrum of adiabatic Gaussian
perturbations which can be well described by a power-law spectrum, tilted from the Harrison--Zel'dovich spectrum, normally tilted so as to provide extra large scale power. The magnitude of the tilt may be modest or pronounced.
The details of structure formation are very sensitive to the matter content of
the universe. It appears that if cold dark matter is the main constituent of
the universe, present observations require that the initial perturbations be
adiabatic --- isocurvature perturbations generate excessively large cmb
anisotropies for the same final density perturbation. Adiabatic perturbations
are exactly what inflation provides. In CDM models, the only remaining
alternative would appear to be texture seeded models, which have been placed
in jeopardy by a combination of microwave anisotropy and velocity data, though
the death blow apparently remains to be struck (Pen, Spergel \& Turok 1992).
%
%
The survey was completed by examining variants on the CDM model which may be
better suited to explaining the observational data. The standard technique is
to utilize additional matter (be it a component of hot dark matter or of a
cosmological constant) to remove short-scale power from the CDM spectrum. Hot
dark matter does this by free-streaming, a cosmological constant by delaying
matter-radiation equality. Because this power can be removed over a much
shorter range of scales than with tilt, it enables an explanation of the
observed deficit of short-scale power relative to intermediate scale power in
the spectrum.

MDM (Mixed dark Model) adds yet another new parameter, roughly speaking an ability
to remove short-scale power from the spectrum while leaving large scales
untouched, and may be necessary should all present observations stand up. It
appears likely that MDM will however need an initial spectrum close to $n=1$
with no gravitational waves if it is to succeed. 
Some studies (Peebles 1984;
Efstathiou et al. 1990a; Turner 1991) has shown that  
lowering the matter density under
the critical value ($ \Omega_m < 1$) and adding a
cosmological constant in order to retain 
a flat Universe ($ \Omega _m + \Omega _\Lambda = 1$), gives good results in the case of $\Omega _m=0.3$. 
Moreover new observational evidences (see Corasaniti, Copeland 2001) indicates that we are living in a $\Lambda \neq 0$ universe.

\begin{acknowledgements}
I would like to thank Dr. P. Valageas for some useful comments. \\
Finally, I would like to thank The Bo$\breve{g}azi$\c{c}i University
Research Foundation for the financial support through the project
code 01B304.
\end{acknowledgements}

\end{document}